\documentclass[11pt]{article}

\usepackage{subfig}
\usepackage{color,colortbl}
\usepackage{epsfig,amssymb,amsfonts,amsmath,graphicx,dsfont,cite,xfrac}
\usepackage{authblk}
\usepackage{rotating}
\usepackage{graphicx}

\usepackage{tikz,pgfplots}
\usetikzlibrary{decorations.pathreplacing}
\usetikzlibrary{shapes.multipart}
\usetikzlibrary{positioning}
\usetikzlibrary{matrix}

\usepackage{cite}
\usepackage[colorlinks=true,linkcolor=blue,citecolor=red]{hyperref}

\title{On the general family of third-order shape-invariant Hamiltonians related to generalized Hermite polynomials}

\author[${1}$]{I. Marquette}
\author[${2}$]{K. Zelaya}

\affil[${1}$]{\footnotesize School of Mathematics and Physics, The University of Queensland, Brisbane, QLD 4072, Australia}

\affil[${2}$]{\footnotesize Nuclear Physics Institute, Czech Academy of Science, 250 68 \v{R}e\v{z}, Czech Republic}

\date{}

\begin{document}

\maketitle

\begin{abstract}
This work reports and classifies the most general construction of rational quantum potentials in terms of the generalized Hermite polynomials. This is achieved by exploiting the intrinsic relation between third-order shape-invariant Hamiltonians and the fourth Painlev\'e equation, such that the generalized Hermite polynomials emerge from the $-1/x$ and $-2x$ hierarchies of rational solutions. Such a relation unequivocally establishes the discrete spectrum structure, which, in general, is composed as the union of a finite- and infinite-dimensional sequence of equidistant eigenvalues separated by a gap. The two indices of the generalized Hermite polynomials determine the dimension of the finite sequence and the gap. Likewise, the complete set of eigensolutions can be decomposed into two disjoint subsets. In this form, the eigensolutions within each set are written as the product of a weight function defined on the real line times a polynomial. These polynomials fulfill a second-order differential equation and are alternatively determined from a three-term recurrence relation (second-order difference equation), the initial conditions of which are also fixed in terms of generalized Hermite polynomials. 
\end{abstract}


\section{Introduction}

The rise of the inverse method~\cite{Nie84} and its relation to the Darboux transformation~\cite{Dar88,Mat91} have paved the way to extend the class of known exactly solvable problems in quantum mechanics. A prime simple yet interesting model was introduced by Mielnik~\cite{Mie84} as a clear example of isospectral models related to the harmonic oscillator.
The latter is true indeed in the context of stationary problems as the Schr\"odinger equation reduces to a Sturm-Liouville problem. Such approach is also known in the literature as SUSY quantum mechanics~\cite{Coo01,Jun96} and equivalently as the factorization method~\cite{Inf51,Mie04,Don07}. When one uses the Darboux transformation on an already known exactly solvable model, one obtains a new model with a potential energy term and eigensolutions constructed in terms of the solutions to a Riccati equation~\cite{Inc56,Sch18}. A more general construction is provided by the so-called Darboux-Crum transformation~\cite{Dar88,Cru55}, which can be seen as consecutive iterations of the conventional Darboux transformation. On the other hand, the class of shape-invariant Hamiltonians~\cite{Gen83,Suk97} has become an useful, yet equivalent, formulation in the study of integrable and solvable models in quantum mechanics~\cite{Mar09a,And00,Sas10,Mar16}. The latter finds particular applications to study new class of orthogonal polynomial solutions in terms of the so-called exceptional polynomials~\cite{Oda10,Sas10,Gom14,Kui15,Gom16,Gom21}. 
The case of higher-order shape-invariant models is also equivalent with the construction of periodic chains of non-linear differential equations associated with non-commuting operators (case $\alpha\neq 0$ in~\cite{Ves93}). This is usually referred to as \textit{dressing chains formalism}~\cite{Ves93}, and provides a mechanism to obtain Backl\"und transformations for some non-linear equations, such as the geometrical approach introduced by Adler~\cite{Adl93}. The latter has been found useful to classify rational solutions for the so-called Painlev\'e systems~\cite{Gom20}.

As mentioned above, when using the Darboux transformation one has to unavoidably deal with a nonlinear problems such as the Riccati equation. But this is the not the only nonlinear equation that emerges from such a formalism. Likewise, Painlev\'e equations have been shown to play a relevant role in the construction of exactly solvable quantum models~\cite{Ber11,Ber16}. This emerges particularly in the context of higher-order shape-invariant models, with the fourth and fifth Painlev\'e equation appearing in third- and fourth-order shape-invariant systems, respectively. Painlev\'e equations are formed by a family of six nonlinear equations defined in terms of complex parameters, whose solutions cannot be expressed in terms of classical functions~\cite{Gro02,Mar06}. In special cases, one can use a particular seed functions to generate a complete hierarchy of solutions through the B\"acklund transformation~\cite{Bas95}, which can be thought as a nonlinear counterpart of the recurrence relations. The fourth and fifth Painlev\'e equations have brought new results in the trend of families of non-orthogonal polynomials generated through non-linear recurrence relations. This includes the hierarchies of rational solutions in terms of the generalized Okamoto, generalized Hermite and Yablonskii-Vorob'ev polynomials, and Umemura polynomials~\cite{Oka86,Cla03,Cla06}. In particular, the fourth Painlev\'e equation emerges quite naturally in third-order shape-invariant SUSY QM~\cite{And00,Mar16,Hus22}, where the behavior of the resulting models is dictated by properly selecting the parameters that define the fourth Painelv\'e equation (usually denoted by $\alpha$ and $\beta$). It is worth to notice that higher-order ladder operators have been alternatively studied in the context of supersymmetric partners for the harmonic oscillator in both the Hermitian~\cite{Fer99,Fer07} and non-Hermitian regimes~\cite{Ros18}.  

In the present work, we focus on the family of third-order shape-invariant Hamiltonian and determine the most general potentials related to the $-1/x$ and $-2x$ hierarchies of rational solutions. The latter is possible since the fourth Painelv\'e equation appears straightforwardly from the third-order shape-invariant condition. In this case, both the potential and the zero-modes (eigensolutions annihilated by the third-order ladder operators) are determined in terms of the generalized Hermite polynomials. This allows establishing the necessary conditions so that the family of potentials becomes free of singularities over the real line. Furthermore, the discrete spectrum of the Hamiltonian can be uniquely identified as well, and it splits into a finite-dimensional and an infinite-dimensional set of equally spaced eigenvalues separated by a gap. Both the dimension of the finite-dimensional set and the gap are determined by the same indexes defining the generalized Hermite polynomials so that we can manipulate both properties of the spectrum at will. This establishes a sharp contrast with the $-2x/3$ hierarchy, where the discrete spectrum is composed of three infinite-dimensional sets of eigenvalues~\cite{Hus22}. This work thus completes the generalization of rational solutions for third-order shape-invariant systems.

The manuscript is structured as follows. In Sec.~\ref{sec:thirdSI}, we briefly discuss the construction of third-order shape invariant Hamiltonians and their relation with the $-1/x$ and $-2x$ hierarchies of rational solutions to the fourth Painlev\'e equation. The conditions for regularity of the potentials are identified, as well as their eigenvalues and zero-modes. In Sec.~\ref{sec:highermodes}, we exploit some of the properties of the ladder operators and use the fact that the zero modes share a common positive-definite function. We use the latter as a weight function, which allows identifying two families of polynomials related to the eigensolutions. Finally, in Sec.~\ref{sec:algebra}, we propose some modified operators fulfilling the $su(2)$ and $su(1,1)$ commutation relations when represented in terms of the appropriate eigensolutions.

\section{Generalized Hermite polynomials and third-order shape-invariant models}
\label{sec:thirdSI}
Among the class of exactly solvable models found in quantum mechanics, one may consider the particular case of shape-invariant Hamiltonians as it provides with an condition to fix the specific form of the Hamiltonian under consideration. According to the original work of Gendenshtein\footnote{Although the term ``shape-invariant'' was not originally coined in~\cite{Gen83}, the main ideas were introduced there.}~\cite{Gen83}, two Hamiltonians $H_{+}$ and $H_{-}$ are said to be shape-invariant if they are related as $H_{+}(\{c_{n}\})=H_{-}(f(\{c_{n}\}))+R(\{c_{n}\})\mathbb{I}$, with $\mathbb{I}$ the identity operator in the corresponding Hilbert space, $\{c_{n}\}$ a set of parameters, and $f(\{c_{n}\})$ and $R(\{c_{n}\})$ functions of the set of parameters. 

We are particularly interested in the class of shape-invariant Hamiltonians $H$ fulfilling the relationship
\begin{equation}
	HA^{\dagger}=A^{\dagger}(H+2\lambda) \, , \quad HA=A(H-2\lambda) \, , \quad \lambda>0 \, ,
	\label{shapeH}
\end{equation}
with $A$ and $A^{\dagger}$  \textit{intertwining operators}. Note that the operators $A^{\dagger}$ and $A$ relate $H_{-}=H$ with $H_{+}=H_{-}+2\lambda$ and $H_{+}=H_{-}-2\lambda$, respectively. Such a construction implies that $A$ and $A^{\dagger}$ are ladder operators in the space of solutions of $H$, where the parameter $\lambda$ defines the energy difference between two contiguous eigenvalues. As $\lambda$ can be absorbed after a suitable change of variables, we consider, without loss of generality, $\lambda=1$.

Let us consider the Hamiltonian $H$ together with its eigenvalue equation
\begin{equation}
H\phi_{n}=E_{n}\phi_{n} \, , \quad H:=-\frac{d^{2}}{dx^{2}}+V(x) \, ,
\label{H}
\end{equation}
with $V(x)$ a real-valued potential to be determined, $E_{n}$ the eigenvalues, and $\phi_{n}(x)\in L^{2}(dx;\mathbb{R})$ the corresponding finite-norm eigensolutions. The norm is defined in terms of the inner product as $\Vert \phi_{n}\Vert^{2}=(\phi_{n},\phi_{n})$, with the inner product
\begin{equation}
(\phi_{m},\phi_{n}):=\int_{\mathbb{R}}dx\,\phi_{m}^{*}(x)\phi_{n}(x)=\delta_{n,m} \, .
\label{inner-1}
\end{equation}
The latter ensures that the finite-norm eigensolutions $\phi_{n}$ are orthogonal, and their respective eigenvalues real.

In general, $A$ and $A^{\dagger}$ may be ladder operators of arbitrary $n$-order. The first-order case corresponds to the harmonic oscillator system, while the second-order case is related to the singular oscillator. In turn, the third-order case brings richer properties when it comes to the specific form of the resulting potential, as there are several families of Hamiltonians related to the shape-invariant condition. Particularly, by adequately factorizing the third-order intertwining operators into lower-order operators, it is possible to realize that the fourth Painlev\'e equation appears quite naturally~\cite{And00}. Several works have been devoted to this particular problem in both the stationary~\cite{Mar16,Can00} and time-dependent case~\cite{Zel20c}. Although finding solutions to the fourth Painlev\'e equation may be a quite challenging task, this has been extensively studied and several families of solutions, usually classified in terms of hierarchies, have been reported in the literature~\cite{Bas95,Cla03,Gro02}. 

The intertwining operators $A$ and $A^{\dagger}$ considered in this work are in general third-order differential operators, which in general can be factorized in terms of lower-order operators. Throughout this work we consider two different, although equivalent, factorizations. We first factorize $A$ and $A^{\dagger}$ through second-order and first-order operators so that the fourth Painlev\'e equation appears quite naturally. Then, a factorization in terms of only first-order operators is introduced to obtain the \textit{zero-modes} of the problem, that is, eigensolutions annihilated by both $A$ and $A^{\dagger}$.

To proceed with the first type of factorization, we consider the approach originally devised by Canata et al.~\cite{Can00} for second-order operators $M$ and $M^{\dagger}$, together with some first-order operators $Q$ and $Q^{\dagger}$ so that
\begin{equation}
A=M^{\dagger}Q \, ,\quad A^{\dagger}=Q^{\dagger}M \, .
\label{Afact1}
\end{equation}
The newly introduced operators are in general given by
\begin{equation}
\begin{aligned}
&M^{\dagger}:=\frac{d^{2}}{dx^{2}}-w(x)\frac{d}{dx}+B(x) \, , \quad &&M=\frac{d}{dx^{2}}+w(x)\frac{d}{dx}+B(x)+w'(x) \, , \\
&Q^{\dagger}:=\frac{d}{dx}+W(x) \, ,\quad &&Q=-\frac{d}{dx}+W(x) \, ,
\end{aligned}
\label{opM}
\end{equation}
with $W(x)$, $B(x)$, and $w(x)$ unknown real valued functions to be determined. The coefficient multiplying the highest derivative in each operator has been fixed, without loss of generality, to one for simplicity\footnote{This coefficient can be considered in general as a function of $x$, however, the term proportional to the second derivative in $H$ is one. This establishes a constraint on the form of the coefficients of the intertwining operators, leading to the forms introduced in~\eqref{opM}.}.

The lower-order operators define by themselves a new set of intertwining relationships 
\begin{align}
& HQ^{\dagger}=Q^{\dagger}(H_{1}+2) \, , && HQ=Q(H_{1}-2) \, ,
\label{intertQ} \\
& HM^{\dagger}=M^{\dagger}H_{1} \, , &&HM=MH_{1} \, .
\label{intertM}
\end{align}
with respect to an auxiliary Hamiltonian $H_{1}=-\frac{d^{2}}{dx^{2}}+V_{1}(x)$, with $V_{1}(x)$ the corresponding potential whose form is not relevant for the ongoing discussion. The latter leads to a differential equation for $w(x)$, and also establishes a relationship for $B(x)$ and $W(x)$ in terms of $w(x)$. See~\cite{Hus22} for more details. After some calculations we get
\begin{equation}
B(x)=\frac{w^{2}}{4}-\frac{w'}{2}-\frac{w''}{2w}+\frac{w'^{2}}{•4w^{2}}+\frac{d}{w^{2}} \, , \quad \quad W(x)=-x-w \, ,
\end{equation}
where $\gamma$ and $d$ are integration constants, together with
\begin{align}
& w''=\frac{(w')^{2}}{2w}+\frac{3}{2}w^{3}+4xw^{2}+2(x^2-\alpha)w+\frac{\beta}{w} \, , 
\label{eqw} \\ 
& \alpha=\gamma+1\, , \quad \beta=2d \, ,
\label{alpha-beta}
\end{align}
from which it s clear that $w(x)$ is a solution of the fourth Painlev\'e equation~\cite{Bas95,Cla03,Olv10}. It is well-known that the parameters $\alpha$ and $\beta$ define the class of solutions of the Painlev\'e equation, defining several hierarchies of solutions depending on the values of such parameters~\cite{Cla08}.

From~\eqref{intertQ}-\eqref{intertM} one also fixes the form of the potential $V(x)$ as a function of $w(x)$ through
\begin{equation}
V(x)=x^{2}-\left(w'-2xw-w^{2}\right)-1 \, .
\label{V}
\end{equation}
Thus, by finding a solution to the fourth Painlev\'e equation, one fixes the form of the potential so that $V(x)$ is a \textit{regular function}, that is, free of singularities, as long as $w(x)$ is a regular solution. Any singularity in $w(x)$ would likely be reflected in the solutions, breaking the finite-norm condition required for the solutions. In this regard, we pay special attention to the properties of $w(x)$ to ensure regularity. Following some previous reports on the construction of third-order shape invariant quantum models~\cite{Mar09,Zel20c}, we particularly focus on the rational solutions of Eq.~\eqref{eqw}. Such a case has been studied for only some specific values of the parameters $\alpha$ and $\beta$, where a connection with two-step Darboux-Crum transformations was found in~\cite{Mar09}, and the relation with the formalism of dressing chains~\cite{Ves93} was also discussed recently in~\cite{Hus22}. 

\subsection{Potentials related to the generalized Hermite polynomials}
\label{sec:pot-GH}
The complete set of rational solutions to the fourth Painlev\'e equation are classified into three hierarchies~\cite{Bas95,Cla03}, namely the ``$-1/x$ hierarchy'', ``$-2x$ hierarchy'', and the ``$-2x/3$ hierarchy''. In~\cite{Hus22}, the $-2x/3$ hierarchy was already explored in full detail by exploiting the fact that the zeros of the conventional Okamoto polynomials are distributed out of the real line. Here, we complement the discussion by considering the remaining two hierarchies of rational solutions, which can be written in terms of generalized Hermite polynomials~\cite{Cla08} $H_{p,q}(x)$. The latter can be computed from the nonlinear recurrence relation
\begin{align}
2pH_{p+1,q}H_{p-1,q}=H_{p,q}H_{p,q}''-(H_{p,q}')^{2}+2pH_{p,q}^{2} \, , 
\label{recurrenceH1}\\
2qH_{p,q+1}H_{p,q-1}=-H_{p,q}H_{p,q}''+(H_{p,q}')^{2}+2qH_{p,q}^{2} \, , 
\label{recurrenceH2}
\end{align}
together with the initial conditions $H_{0,0}=H_{1,0}=H_{0,1}=1$ and $H_{1,1}=2x$. These initial conditions reveal that $H_{0,q}=H_{p,0}=1$ for $p,q\in\mathbb{Z}^{+}$. See Tab.~\ref{tab1} for the explicit form of some polynomials $H_{p,q}$.

\begin{table}
	\centering
	\footnotesize{\begin{tabular}{ccp{10cm}}
			\hline
			$p$ & $q$ & $H_{p,q}(x)$ \\
			\hline
			1 &  1 & $2x$ \\
			2 &  1 & $4 x^2-2$ \\
			3 &  1 & $8 x^3-12 x$  \\
			1 &  2 & $4 x^2+2$ \\
			2 &  2 & $16 x^4+12$ \\
			3 &  2 & $64 x^6-96 x^4+144 x^2+72 $\\
			1 &  3 & $8 x^3+12 x$\\
			2 &  3 & $64 x^6+96 x^4+144 x^2-72$\\
			3 &  3 & $512 x^9+2304 x^5-4320 x$\\
			1 &  4 & $16 x^4+48 x^2+12$\\
			2 &  4 & $256 x^8+1024 x^6+1920 x^4+720$\\
			3 &  4 & $4096 x^{12}+12288 x^{10}+46080 x^8+30720 x^6-57600 x^4+172800 x^2+43200$\\
			\hline
	\end{tabular}}
	\caption{Generalized Hermite polynomials $H_{p,q}(x)$ computed from~\eqref{recurrenceH2} and the initial conditions $H_{0,0}=H_{0,1}=1$ and $H_{1,1}=2x$. The case $H_{p,0}=H_{0,q}=1$ has been excluded from the table.}
	\label{tab1}
\end{table}

Alternatively, the generalized Hermite polynomials have a Wronskian representation in terms of Hermite polynomials of the form~\cite{Cla08} 
\begin{equation}
H_{p,q}(x)\propto Wr(H_{p}(x),\ldots,H_{p+q-1}(x)) \, .
\label{wronskian}
\end{equation}
This is particularly useful as zeros distributions for Wronskian determinant of orthogonal polynomials have been extensively studied~\cite{Kar60,Adl94,Fel12,Gar15}. In particular, in~\cite{Kar60} the authors proved two theorems regarding the number of zeros for Wronskians composed of contiguous orthogonal polynomials. The zeros are distributed distributed over the domain of the corresponding weight function. Such theorems applies straightforwardly to the Wronskian~\eqref{wronskian}, from which we conclude that $H_{q,p}(x)$ has no zeros in $x\in\mathbb{R}$ for even $q$, and $p$ simple zeros for odd $q$.

Rational solutions in the $-1/x$ and $-2x$ hierarchies are divided in three classes, the specific form of which is dictated by the values of $\alpha$ and $\beta$, and given by~\cite{Bas95,Mar06}
\begin{align}
&w^{[1]}_{p,q}(x)=\frac{d}{dx}\ln\frac{H_{p+1,q}}{H_{p,q}} \, , \quad &&\alpha=2p+q+1 \, , \quad &&\beta=-2q^{2} \, ,
\label{w1}\\
&w^{[2]}_{p,q}(x)=\frac{d}{dx}\ln\frac{H_{p,q}}{H_{p,q+1}} \, , \quad &&\alpha=-(p+2q+1) \, , \quad &&\beta=-2p^{2} \, ,
\label{w2}\\
&w^{[3]}_{p,q}(x)=-2x+\frac{d}{dx}\ln\frac{H_{p,q+1}}{H_{p+1,q}} \, , \quad &&\alpha=q-p \, , \quad &&\beta=-2\left(p+q+1\right)^{2} \, ,
\label{w3}
\end{align}
with $q$ and $p$ integers. 

These rational solutions may be rewritten in an alternative, and sometimes convenient, form utilizing the appropriate B\"acklund transformations (see also App.~A in~\cite{Hus22} for details). This leads to
\begin{equation}
w^{[1]}_{p,q}(x)=2q\frac{H_{p+1,q-1}H_{p,q+1}}{H_{p,q}H_{p+1,q}}, \quad 
w^{[2]}_{p,q}(x)=-2p \frac{H_{p+1,q}H_{p-1,q+1}}{H_{p,q+1}H_{p,q}}, \quad
w^{[3]}_{p,q}(x)=-\frac{H_{p+1,q+1}H_{p,q}}{H_{p+1,q}H_{p,q+1}} ,
\label{wr}
\end{equation}
and we use henceforth both the representations~\eqref{w1}-\eqref{w3} and~\eqref{wr} interchangeably as per required.

Notice that the rational solutions shown in~\eqref{w1}-\eqref{w3} are free of singularities for $x\in\mathbb{R}$ if the respective generalized Hermite polynomials $H_{p,q}(x)$ are nodeless functions. As we discussed above, this is the case for even $q$. Thus, regular solutions $w(x)$ for the $-1/x$ and $-2x$ hierarchies are obtained only in the following case: 
\begin{equation}
w(x) \equiv w^{[1]}_{p,2q}(x)=\frac{d}{dx}\ln\frac{H_{p+1,2q}}{H_{p,2q}} \, , \quad \alpha_{p,q}=2p+2q+1 \, , \quad \beta_{p,q}=-8q^{2} \, ,
\label{wk}
\end{equation}
for $q,p=0,1,\cdots$. The latter fixes the integration constants in~\eqref{alpha-beta} as $d=-4q^{2}$ and $\gamma=2p+2q$.

The potential $V(x)$ in~\eqref{V} is then free of singularities and identified explicitly in terms of the generalized Hermite polynomials. This can be further simplified utilizing the B\"acklund and Schlesinger transformations~\cite{Cla08} for the term $w'-(2xw+w^2)$. It is useful to provide a simplified expression for $w'+(2xw+w^2)$ as well, as it will be useful in the sequel. After some calculations (see also~\cite{Hus22} and App.~A therein) we get
\begin{align}
& w'-(2xw+w^2)=-8q\frac{H_{p+1,2q+1}H_{p+1,2q-1}}{(H_{p+1,2q})^{2}}+4q=2\frac{d^{2}}{dx^{2}}\ln H_{p+1,2q}-4q \, , 
\label{wminus}\\
& w'+(2x+w^2)=8q\frac{H_{p,2q+1}H_{p,2q-1}}{(H_{p,2q})^{2}}-4q=-2\frac{d^{2}}{dx^{2}}\ln H_{p,2q}+4q \, ,
\label{wplus}
\end{align}
which in turn leads to the potential
\begin{equation}
V_{p,q}(x)\equiv V(x)=x^{2}+4q-1-\frac{d^{2}}{dx^{2}}\ln H_{p+1,2q} \, .
\label{potVk}
\end{equation}
In this form, the third-order shape-invariant Hamiltonian and its family of regular potential $V_{p,q}(x)$ have been identified in the most general case for the $-1/x$ and $-2x$ hierarchies of rational solutions of the fourth Painlev\'e equation. The Hamiltonian related to the $-2x/3$ hierarchy has been already constructed in~\cite{Hus22}, and will not be considered throughout this work. 

\subsection{Zero-modes associated with $H$}
\label{sec:spectrum}
The procedure to determine the set of eigensolutions is quite straightforward. First, one may recall that $A$ and $A^{\dagger}$ are annihilation and creation operators on the eigensolutions of $H$, which is an immediate consequence of the shape-invariant condition~\eqref{shapeH}. We can then compute those eigensolutions annihilated by $A$ and their corresponding eigenvalues, the rest of the eigenfunctions are generated by the iterated action of the creation operator $A^{\dagger}$ and the corresponding eigenvalues are increased by two units after each iteration. Additional caution must be taken in this case, as there might be finite-norm eigensolutions annihilated by $A^{\dagger}$ that truncate the aforementioned iteration. This would lead to a finite-dimensional sequence of solutions. The set of solutions annihilated by $A$ and $A^{\dagger}$ are henceforth called \textit{zero-modes}. Since $A$ and $A^{\dagger}$ are third-order differential operators, there are six annihilated solutions in total, three for each operator. However, not all of such solutions have a finite norm, and this has to be studied once we have the explicit form of the zero modes.

As mentioned in the previous section, the third-order intertwining operators $A$ and $A^{\dagger}$ have been factorized in terms of the second-order and first-order operators introduced in~\eqref{Afact1} to show the appearance of the fourth Painlev\'e equation explicitly. In order to determine the eigensolutions of $H$, it is more convenient and simple if we further decompose the second-order operators as $M^{\dagger}=M_{1}^{\dagger}M_{2}^{\dagger}$ and $M=M_{2}M_{1}$. The first-order operators $M_{1}^{\dagger}$ and $M_{2}^{\dagger}$ are of the form
\begin{equation}
M^{\dagger}_{1}:=\frac{d}{dx}+W_{1}(x) \, , \quad M^{\dagger}_{2}:=\frac{d}{dx}+W_{2}(x) \, ,
\label{M1M2D}
\end{equation}
where the real-valued functions $W_{1,2}(x)$ are computed from the factorization $M^{\dagger}=M_{1}^{\dagger}M_{2}^{\dagger}$ after comparing term by term with~\eqref{opM}. The straightforward calculations lead to~\cite{Mar16}
\begin{equation}
W_{1}(x)=\frac{(w'-2xw-w^2) \pm 4q}{2w}+x \, , \quad  W_{2}(x)=-\frac{(w'+2xw+w^2) \pm 4q}{2w}+x \, ,
\label{WW}
\end{equation}
where the sign $\pm$ can be fixed, without loss of generality, to ``$+$.'' 

The terms inside parenthesis in~\eqref{WW} is the same as the one that appeared in the potential $V(x)$, which can be simplified by means of B\"acklund transformations, as we did in the previous section. We thus get
\begin{equation}
W_{1}=x+\frac{d}{dx}\ln\frac{H_{p+1,2q-1}}{H_{p+1,2q}} \, , \quad W_{2}=-x+\frac{d}{dx}\ln\frac{H_{p,2q}}{H_{p+1,2q-1}} \, . 
\label{W1W2}
\end{equation}
together with
\begin{equation}
W=-(x+w)=-x+\frac{d}{dx}\ln\frac{H_{p,2q}}{H_{p+1,2q}} \, .
\label{W}
\end{equation}

The latter handy expressions allow us obtaining explicit expressions for the eigensolutions. To see this, we first solve
\begin{equation}
A\phi_{0;k}=M^{\dagger}Q\phi_{0;k}=M_{1}^{\dagger}M_{2}^{\dagger}Q\phi_{0;k}=0 \, , \quad k\in\{1,2,3\} \, ,
\label{kernelA}
\end{equation}
from which we obtain three zero-modes $\phi_{0;k}$. The zero-mode $\phi_{0;1}$ is particularly simple, as it is just that solution of $Q\phi_{0;1}=0$, leading to $\phi_{0;1}(x)\propto\mathcal{N}_{0;1}e^{\int dx W(x)}$. The other two zero-modes require more elaborated calculations, which has been previously done in~\cite{Mar16}, and recently by exploiting the first-order operators in~\cite{Hus22}. Here we just take the final result, which leads to the zero-modes of the form $\phi_{0;2}(x)\propto\mathcal{N}_{0;2}\left(W-W_{2}\right)e^{-\int dx W^{\pm}_{2}(x)}$ and $\phi_{0;3}(x)\propto\mathcal{N}_{0;3}\left(2\sqrt{-d}+(W-W_{2})(W_{1}+W_{2}) \right)e^{-\int dx W^{\pm}_{1}(x)}$. The corresponding eigenvalues~\cite{Hus22} are given by $E_{0;1}=0$, $E_{0;2}=\gamma+2-\sqrt{-d}$, and $E^{\pm}_{0;3}=\gamma+2+\sqrt{-d}$.

\begin{equation}
\begin{aligned}
&\phi_{0;1}(x)=\frac{1}{\mathcal{N}_{0;1}}\left(\frac{e^{-\frac{x^{2}}{2}}}{H_{p+1,2q}}\right)H_{p,2q} \, , \quad \phi_{0;2}(x)=\frac{1}{\mathcal{N}_{0;2}}\left(\frac{e^{\frac{x^{2}}{2}}}{H_{p+1,2q}} \right)H_{p+2,2q-1} \, , \\
&\phi_{0;3}(x)=\frac{1}{\mathcal{N}_{0;3}}\left(\frac{e^{-\frac{x^{2}}{2}}}{H_{p+1,2q}} \right)H_{p+1,2q+1} \, ,
\end{aligned}
\end{equation}
with the respective eigenvalues $E_{0;1}=0$, $E_{0;2}=2p+2$, and $E_{0;3}=2p+4q+2$.

Since the polynomials $H_{p,2q}(x)$ are nodeless in $x\in\mathbb{R}$, the zero-mode $\phi_{0;1}(x)$ is both a regular and nodeless function. Thus, $\phi_{0;1}(x)$ is the ground eigensolution of $H$ with $E_{0;1}=0$. From the asymptotic behavior of the other two eigensolutions, one may realize that $\phi_{0,2}(x)$ diverges asymptotically. We thus have $\phi_{0;2}(x)\not\in L^{2}(dx;\mathbb{R})$ and $\phi_{0;3}(x)\in L^{2}(dx;\mathbb{R})$. 

On the other hand, the zero-modes $\Phi_{0;j}$, for $j=1,2,3$, annihilated from the creation operator $A^{\dagger}$, i.e., $A^{\dagger}\Phi_{0;j}=0$, are computed in a similar way~\cite{Zel20c}. After several calculations we get
\begin{align}
& \Phi_{0;1}(x)=\frac{1}{\widetilde{\mathcal{N}}_{0;1}}\left(\frac{e^{\frac{x^{2}}{2}}}{H_{p+1,2q}} \right)H_{p+1,2q-1} \, , \quad \Phi_{0;2}(x)=\frac{1}{\widetilde{\mathcal{N}}_{0;2}}\left(\frac{e^{-\frac{x^{2}}{2}}}{H_{p+1,2q}} \right)H_{p,2q+1} \, , \\ 
& \Phi_{0;3}(x)=\frac{1}{\widetilde{\mathcal{N}}_{0;3}}\left(\frac{e^{\frac{x^{2}}{2}}}{H_{p+1,2q}} \right)H_{p+1,2q+1} \, 
\label{phiAdagger}
\end{align}
where the corresponding eigenvalues are in this case $\widetilde{E}_{0;1}=2p+4q$, $\widetilde{E}_{0;2}=2p$, and $\widetilde{E}_{0;3}=-2$. 

In this case, we may note that $\Phi_{0;2}(x)\in L^{2}(dx;\mathbb{R})$ is the only finite-norm eigensolution. Only one of the two sequences is truncated by the action of $A^{\dagger}$, and we thus get one finite-dimensional sequence. The other sequence begins with $\phi_{0;3}(x)$ and the eigenvalue $E_{0;3}=2p+4q+2$, where the action of $A^{\dagger}$ creates eigensolutions with no upper bound. We thus have an infinite-dimensional sequence in this case.

From the latter considerations, we select the finite-norm solutions and their respective eigenvalues, which we classify in two sequence labeled by $\phi^{(p,q)}_{n;1}(x)$ and $\phi^{(p,q)}_{n;2}(x)$ for the finite-dimensional and infinite-dimensional sequence of eigensolutions, respectively. In this form, we can summarize the relevant zero-modes as
\begin{align}
& \phi^{(p,q)}_{n=0;1}(x)\equiv\phi_{0;1}(x)=\frac{1}{\mathcal{N}_{0;1}^{(p,q)}}\left(\frac{e^{-\frac{x^{2}}{2}}}{H_{p+1,2q}} \right)H_{p,2q} \, , \quad &&E_{0;0}=0 \, , \\
& \phi^{(p,q)}_{n=p;1}(x)\equiv\Phi_{0;2}(x)=\frac{1}{\mathcal{N}_{p;1}^{(p,q)}}\left(\frac{e^{-\frac{x^{2}}{2}}}{H_{p+1,2q}} \right)H_{p,2q+1} \, , \quad &&E_{p;1}=2p \, , \\
& \phi^{(p,q)}_{n=0;2}(x)\equiv\phi_{0;3}(x)=\frac{1}{\mathcal{N}_{0;2}^{(p,q)}}\left(\frac{e^{-\frac{x^{2}}{2}}}{H_{p+1,2q}} \right)H_{p+1,2q+1} \, , \quad &&E_{0;2}=2p+4q+2 \, ,
\label{zero-final}
\end{align}
with $\mathcal{N}^{(p,q)}_{n;j}$, for $j=1,2$, the associated normalization constants for each sequence. Since $H_{p,2q}(x)$ is free of zeros for $x\in\mathbb{R}$, the zero-mode $\phi_{0;1}^{(p,q)}(x)$ is nodeless and thus the ground state solution of $H$. This coincides with the fact that $E_{0,0}=0$ is the lowest eigenvalues of $H$. Similarly, from the distribution of real zeros of $H_{p,q}$ discussed after Eq.~\eqref{wronskian}, we conclude that $\phi_{p;1}^{(p,q)}(x)$ and $\phi_{0;2}^{(p,q)}(x)$ have exactly $p$ and $p+1$ real simple zeros, respectively. 



\section{Spectral information}
\label{sec:highermodes}
It is now clear that $A^{\dagger}$ acts on $\phi^{(p,q)}_{0;1}(x)$ only $p$ times, and any further iteration annihilates the resulting eigensolution. In turn, the action on $\phi^{(p,1)}_{0;2}(x)$ may be performed indefinitely. Given that the eigenvalue is increased by two units after each iteration, we can determine the remaining eigenvalues associated to each sequence, which are simply given by
\begin{align}
	& E_{n;1}=2n \, , \quad && n=0,\ldots,p \, , 
	\label{En1}\\
	& E_{n;2}=2n+2p+4q+2 \, , \quad && n=0,1,\ldots \, ,
	\label{En2}
\end{align}
such that the eigenvalues~\eqref{zero-final} are recovered for the appropriate values of $n$.

Since the creation operator $A^{\dagger}$ is represented by a third-order differential operator, any attempt to determine closed expressions for the eigensolutions might become a quite challenging task. We thus resort to an alternative way to construct such eigensolutions by means of recurrence relations. To this end, it is worth noticing that every zero-mode $\phi^{(p,q)}_{0,j}$ has a common factor of the form $e^{-\frac{x^{2}}{2}}/H_{p+1,2q}(x)$, which is independent of the sequence index $j$. We can exploit the latter and conveniently rewrite the eigenfunctions $\phi^{(p,q)}_{n;j}(x)$ as
\begin{equation}
\phi^{(p,q)}_{n;j}(x)=\frac{\mu^{(p,q)}(x)}{\mathcal{N}^{(p,q)}_{n,j}}P_{n;j}^{(p,q)}(x) \, , \quad \mu^{(p,q)}(x):=\frac{e^{-\frac{x^{2}}{2}}}{H_{p+1,2q}(x)} \, , \quad j\in\{1,2\} \, ,
\label{phin}
\end{equation}
where $\mu^{(p,q)}(x)$ has been introduced as a weight function, and now it is left to determine the unknown function $P_{n;j}^{(k)}(x)$, which must be constrained to fulfill
\begin{equation}
P_{0;1}^{(p,q)}(x):=H_{p,2q}(x) \, , \quad P_{0;2}^{(p,q)}(x):=H_{p+1,2q+1}(x) \, .
\label{initial}
\end{equation}
The latter allows recovering the zero-modes $\phi_{0;j}(x)$ in the appropriate limit. Eq.~\eqref{initial} can be thought as a set of initial conditions, which, as discussed below, serves as a departing point to compute any other solution.

Interestingly, the normalization constants $\mathcal{N}^{(p,q)}_{n;j}(x)$ can be expressed in terms of the one associated with the zero modes $\mathcal{N}^{(p,q)}_{0;j}$, for $j=1,2$. This can be done even without knowing the explicit form of $P_{n;j}^{(p,q)}(x)$. To this end, let us consider the action of the ladder operators $A$ and $A^{\dagger}$ on any arbitrary element of $\mathfrak{h}_{j}$. Given that the ladder operators are mutually adjoint, we obtain 
\begin{equation}
	A^{\dagger}\phi^{(p,q)}_{n;j}=C^{(p,q)}_{n+1;j}\phi_{n+1;j}^{(p,q)} \, , \quad A\phi_{n;j}^{(p,q)}=C^{(p,q)}_{n;j}\phi_{n-1;j}^{(p,q)} \, ,  
	\label{Adagger}
\end{equation}
where $C_{n;j}$ is a proportionality constant computed from $(\phi_{n;j},AA^{\dagger}\phi_{n;j})=\vert C_{n+1;j}\vert^{2}$, where the term $AA^{\dagger}$ can be decomposed by means of the previously introduced first-order operators. This leads to a polynomial of degree three for $H$ (see~\cite{Hus22} for further details) of the form $AA^{\dagger}=M_{1}^{\dagger}M_{2}^{\dagger}QQ^{\dagger}M_{2}M_{1}=(H+2)(H-\epsilon_{1})(H-\epsilon_{2})$ with $\epsilon_{1}=2p+4q$ and $\epsilon_{2}=2p$. In this way, the proportionality constants become
\begin{align}
	& C^{(p,q)}_{n+1;1}:=\sqrt{2^{3}(n+1)\left(p+2q-n\right)\left(p-n\right)} \, , \quad &&n=0,\ldots,p-1 \, , 
		\label{C1}\\
	& C^{(p,q)}_{n+1;2}:=\sqrt{2^{3}(n+1)\left(n+2q+1\right)\left(n+p+2q+2\right)} \, , \quad &&n=0,1,\ldots \, ,
	\label{C2}
\end{align}
where $C^{(p,q)}_{p+1;1}=0$ and $C^{(p,q)}_{0;1}=C^{(p,q)}_{0;2}=0$, as expected as the corresponding eigensolutions are annihilated.

From the latter, together with Eq.~\eqref{Adagger}, the normalization factors $\mathcal{N}_{n;j}$, for $j=1,2$, may be determined in terms of $\mathcal{N}_{0;j}$ through $\mathcal{N}_{n+1;j}=\mathcal{N}_{0;j}\prod_{k=0}^{n}C_{k+1;j}$. We thus get the corresponding normalization factors for each mode as
\begin{align}
& \mathcal{N}^{(p,q)}_{n;1}=\mathcal{N}^{(p,q)}_{0;1}\sqrt{2^{3n}n!(-p)_{n}(-p-2q)_{n}} \, , \quad && n=0,\ldots,p-1 \, ,
\label{N01}\\
& \mathcal{N}^{(p,q)}_{n;2}=\mathcal{N}^{(p,q)}_{0;2}\sqrt{2^{3n}n!(2q+1)_{n}(2q+p+2)_{n}} \, , \quad && n=0,1,\ldots \, .
\label{N02}
\end{align}

\begin{figure}[t]
	\centering
	\begin{tikzpicture}
		\draw (1,9.5) node {$\mathcal{H}$};
		\draw[black,dotted,very thick] (2,0)--(0,0);
		\draw[blue,line width=1mm] (2,1)--(0,1) node[at start,anchor=west,text=black] {$0$} node[at end,anchor=east,text=black] {};
		\draw[blue,very thick] (2,2)--(0,2) node[at start,anchor=west,text=black] {$2$} node[at end,anchor=east,text=black] {};
		
		\filldraw[black] (1,19/6) circle (1pt);
		\filldraw[black] (1,18/6) circle (1pt);
		\filldraw[black] (1,17/6) circle (1pt);
		
		\draw[blue,very thick] (2,4)--(0,4) node[at start,anchor=west,text=black] {$2(p-1)$} node[at end,anchor=east,text=black] {};
		\draw[blue,line width=1mm] (2,5)--(0,5) node[at start,anchor=west,text=black] {$2p$} node[at end,anchor=east,text=black] {};
		
		\filldraw[black] (1,37/6) circle (1pt);
		\filldraw[black] (1,36/6) circle (1pt);
		\filldraw[black] (1,35/6) circle (1pt);
		
		\draw[color=green!50!black,line width=1mm] (2,7)--(0,7) node[at start,anchor=west,text=black] {$2(p+2q+1)$} node[at end,anchor=east,text=black] {};
		\draw[color=green!50!black,very thick] (2,8)--(0,8) node[at start,anchor=west,text=black] {$2(p+2q+2)$} node[at end,anchor=east,text=black] {};
		
		\filldraw[black] (1,55/6) circle (1pt);
		\filldraw[black] (1,54/6) circle (1pt);
		\filldraw[black] (1,53/6) circle (1pt);
		
		\draw[red,dashed,very thick,->] (1,1)--(1,0) node[midway,anchor=west] {$A$};
		\draw[red,dashed,very thick,->] (0,7)--(-2,7)--(-2,0) node[midway,anchor=west] {$A$};
		\draw[red,dashed,very thick,->] (0,5)--(-1,5)--(-1,0) node[midway,anchor=west] {$A^{\dagger}$};
		\draw[black,very thick,->] (1,1)--(1,2) node[midway,anchor=west] {$A^{\dagger}$};
		\draw[black,very thick,-] (1,2)--(1,2.6) node[midway,anchor=west] {};
		\draw[black,very thick,->] (1,3.4)--(1,4) node[midway,anchor=west] {};
		\draw[black,very thick,->] (1,4)--(1,5) node[midway,anchor=west] {};
		
		\draw[black,very thick,->] (1,7)--(1,8) node[midway,anchor=west] {$A^{\dagger}$};
		\draw[black,very thick,->] (1,8)--(1,8.5) node[midway,anchor=west] {};
		
		
		\draw [decorate,decoration={brace,amplitude=10pt,mirror,raise=4pt},yshift=0pt]
		(4.5,1.2) -- (4.5,4.8) node [black,midway,xshift=2cm] {\footnotesize{
				$(p+1)$-dim. seq.}};
		
		\draw [decorate,decoration={brace,amplitude=10pt,mirror,raise=4pt},yshift=0pt]
		(4.5,6.8) -- (4.5,9.2) node [black,midway,xshift=2cm] {\footnotesize{
				inf.-dim. seq.}};
		
		\draw [decorate,decoration={brace,amplitude=10pt,mirror,raise=4pt},yshift=0pt]
		(4.5,5.4) -- (4.5,6.6) node [black,midway,xshift=2cm] {\footnotesize{
				$2q$-dim. gap}};
		
	\end{tikzpicture}
	\label{fig:FGenH}
	\caption{\footnotesize{(Color online) Gapped spectrum associated with the $-1/x$ and $-2x$ hierarchies of rational solutions to the fourth Painlev\'e equation.}}
	\label{fig:spectrum}
\end{figure}
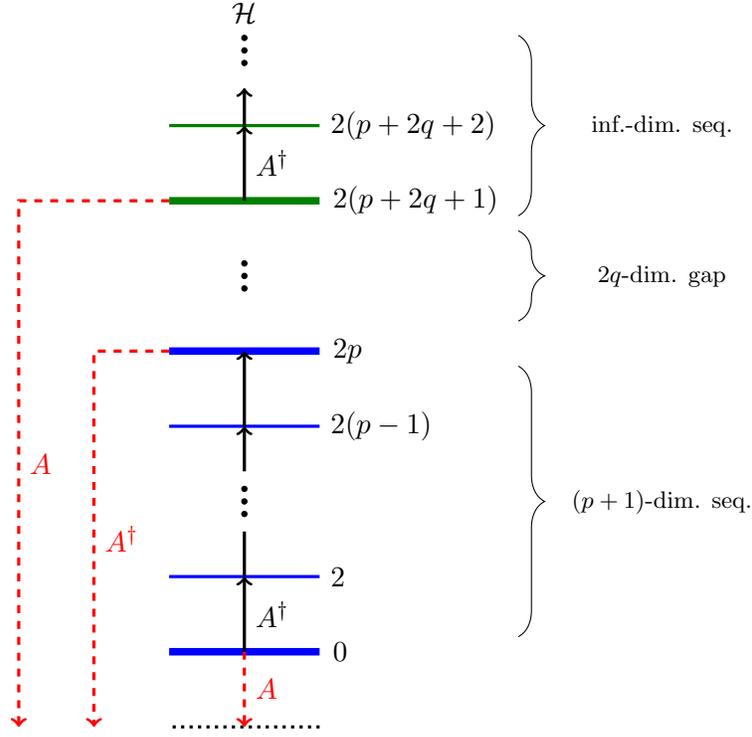

\subsection{Differential equation for $P_{n;j}^{(p.1)}(x)$}
\label{sec:diff}
Now, we can determine the differential equation associated to the functions $P^{(p,q)}_{n;j}(x)$. Since we have two sequence of solutions, it is convenient to decompose the space of solutions of $H$, denoted by $\mathcal{H}$, as the direct sum of the subspaces generated from each sequence. That is,
\begin{equation}
\mathcal{H}=\mathfrak{h}_{1}\oplus\mathfrak{h}_{2} \, , \quad \mathfrak{h}_{1}=\operatorname{Span}\{\phi_{n;1}\}_{n=0}^{p} \, , \quad \mathfrak{h}_{2}=\operatorname{Span}\{\phi_{n;2}\}_{n=0}^{\infty} \, ,
\label{subspace}
\end{equation}
where the ladder operators $A$ and $A^{\dagger}$ act as endomorphisms on the elements of the subspaces $\mathfrak{h}_{j}$, that is, $A:\mathfrak{h}_{j}\rightarrow\mathfrak{h}_{j}$ and $A^{\dagger}:\mathfrak{h}_{j}\rightarrow\mathfrak{h}_{j}$ for $j\in\{1,2\}$, see Fig.~\ref{fig:spectrum}. 

By substituting~\eqref{phin} into the eigenvalue problem~\eqref{H}, with $V(x)$ given in~\eqref{potVk}, and using the identities
\begin{equation}
\begin{aligned}
&\frac{1}{\mu^{(p,q)}}\frac{d^{2}\mu^{(p,q)}}{dx^{2}}:=-1-\frac{d^{2}}{dx^{2}}\ln H_{p+1,2q}+\left(x+\frac{d}{dx}\ln H_{p+1,2q}\right)^{2} \, , \\
&\frac{1}{\mu^{(p,q)}}\frac{d\mu^{(p,q)}}{dx}:=-x-\frac{d}{dx}\ln H_{p+1,2q} \, .
\end{aligned}
\label{mu-der}
\end{equation}
one obtains a second-order differential equation for $P^{(p,q)}_{n;j}$ of the form
\begin{equation}
\frac{d^{2}P_{n;j}^{(p,q)}}{dx^{2}}-2\left(x+\frac{H'_{p+1,2q}}{H_{p+1,2q}} \right)\frac{dP_{n;j}^{(p,q)}}{dx}+\left( \frac{H_{p+1,2q}''}{H_{p+1,2q}} + 2x \frac{H_{p+1,2q}'}{H_{p+1,2q}}+E_{n;j}-4q \right)P_{n;j}^{(p,q)}=0 \, .
\label{diff-Pn}
\end{equation}

Although obtaining the explicit form out of the differential equation might be even more difficult than using the action of the creation operator on the zero-modes, the latter provides us with essential information to be exploited later in the construction of recurrence relations, which are much easier to solve by iterative means. 

In particular, one can obtain information about the behavior of $P^{(p,q)}_{n;j}(z)$, for $z\in\mathbb{C}$, around the complex zeros of $H_{p+1,2q}(z)$, which are singularities in the coefficients of the differential equation. The behavior of the solutions around those points will be useful in the next section. 

To begin with, let us rewrite the generalized Hermite polynomials as $H_{p+1,2q}(z)=\prod_{k=1}^{pq}(z-z_{k})$, where $z_{k}$ are the respective zeros, whose specific value is no relevant here. The latter expression is valid since the zeros of the generalized Hermite polynomials are all simple~\cite{Cla08}. Next, we expand the solution around one of the singularities, let say $z_{s}$, in power series of the form $P_{n;j}^{(p,q)}(z)=\sum_{m}c_{m}(z-{z_{s}})^{m+\ell}$, with $\ell$ to be determined from the secular equation. It is well known that if $\ell$ is a positive integer or zero, the solution has no pole at such point, neither simple nor essential. By substituting the series expansion and the product expansion of $H_{p+1,2q}(z)$ into~\eqref{diff-Pn}, one arrives at the secular equation $\ell(\ell-3)=0$, which is the same regardless of the singularity $z_{s}$ choose in the expansion. Clearly, $\ell=0$ or $\ell=3$, and thus the solution has no poles at the singularities. Thus, $P_{n;j}^{(p,q)}$ are analytic functions across the complex plane (entire functions).

\subsection{Three-term recurrence relation for $P_{n;j}^{(p,q)}(x)$}
We can now show one of the main results of this work. In analogy to classical polynomials~\cite{Sze59,Chi78}, it is possible to find a three-term recurrence relation (second-order finite-difference equation) in order to compute $P_{n;j}^{(p,1)}(x)$. The latter is computationally simpler to implement than the differential equation or the direct application of the $(A^{\dagger})^{n}$ on the zero-modes, for arbitrary $n$. To this end, let compute the explicit action of $A$ and $A^{\dagger}$ on $\phi_{n;j}^{(p,q)}$ given in~\eqref{phin} so that, with the aid of~\eqref{diff-Pn}, we determine a relationship among $P_{n;j}(x)$, its derivative $P_{n;j}'(x)$, and contiguous functions $P_{n\pm 1;j}(x)$. 

From the operators $A$ and $A^{\dagger}$, and after some calculations, we obtain the relations
\begin{equation}
\mathfrak{R}_{n;j}^{(p,q)}\frac{d P_{n}^{(p,q)}}{dx}=
\left( \mathfrak{R}_{n;j}^{(p,q)}\frac{d}{dx}\ln H_{p,2q}+E_{n;j}w_{p,2q}^{[1]}\right)P^{(p,q)}_{n;j}+C_{n;j}^{(p,1)}\frac{\mathcal{N}_{n;j}^{(p,q)}}{\mathcal{N}_{n-1;j}^{(p,q)}}P_{n-1;j}^{(p,q)} \, ,
\end{equation}
and
\begin{multline}
(\mathfrak{R}_{n;j}^{(p,q)}+2)\frac{d P_{n}^{(p,q)}}{dx}=\\
-\left( \mathfrak{R} _{n;j}^{(p,q)}\left(2x+\frac{d}{dx}\ln H_{p+1,2q-1}\right)+(2(p+2q)+E_{n;j})w_{p+1,2q-1}^{[2]}\right) P_{n;j}^{(p,q)} \\
-C_{n+1;j}^{(p,q)}\frac{\mathcal{N}_{n;j}^{(p,q)}}{\mathcal{N}_{n+1;j}^{(p,q)}}P_{n+1;j}^{(p,q)} \, ,
\end{multline}
respectively, where $C_{n;j}^{(p,q)}$ are given in~\eqref{C1}-\eqref{C2}, and the normalization constants $\mathcal{N}_{n;j}^{(p,q)}$ in~\eqref{N01}-\eqref{N02}. On the other hand, we have
\begin{equation}
\mathfrak{R}_{n;j}^{(p,q)}=E_{n;j}-2(p+1)-w_{p,2q}^{[1]}w_{p+1,2q-1}^{[2]} \, ,
\end{equation}
with $w_{p,q}^{[k]}$, for $k=1,2,3$, defined in Sec.~\ref{sec:pot-GH}.

The latter expressions are useful as we can vanish the term proportional to the derivative of $P_{n;j}^{(p,q)}(x)$, leading to the three-term recurrence relation 
\begin{multline}
P_{n+1;j}^{(p,q)} + (C^{(p,q)}_{n;j})^{2}\left( \frac{\mathfrak{R}_{n;j}^{(p,q)}+2}{\mathfrak{R}_{n;j}^{(p,q)}}\right) P_{n-1;j}^{(p,q)} \\
+\left( (\mathfrak{R}_{n;j}^{(p,q)}+2)w_{p,2q-1}^{[3]}+(2(p+2q)-E_{n;j})w_{p+1,2q-1}^{[2]}+E_{n;j}\frac{\mathfrak{R}_{n;j}^{(p,q)}+2}{\mathfrak{R}_{n;j}^{(p,q)}} w_{p,2q}^{[1]} \right)P_{n;j}^{(p,q)} = 0 \, , 
\label{Pn}
\end{multline}
valid for $n=1,\ldots,p-1$ for $j=1$, and $n=1,\ldots$ for $j=2$. 

It is worth remarking that~\eqref{Pn} is a second-order finite-difference equation, and as such, it is necessary to specify two initial conditions to uniquely determine the complete set of solutions. In this regard, $P_{0;j}^{(p,q)}(x)$ in~\eqref{initial} servers as the first initial condition. Since $C_{0;j}^{(p,q)}=0$ for $j=1,2$, we can fix $P_{-1;j}^{(p,q)}(x)=0$ as the second initial condition, as it is customary for classical orthogonal polynomials.

From~\eqref{Pn} one might conclude that $P_{n;j}^{(p,q)}(x)$ is a rational solution, but the analysis of singularities discussed in Sec.~\ref{sec:diff} allows us to get more information. Let us first extend the domain into the complex plane $z\in\mathbb{C}$, i.e., $P_{n;j}^{(p,q)}(z)$. Even though we are interested in solutions over the real line $x\in\mathbb{R}$, by extending it into the complex plane one can see the distribution of poles, which in turn shows the behavior of the solutions on the real line. To see this, we first assume that the solutions are indeed rational so that they can be written as $P_{n;j}^{(p,q)}(z)=f_{n;j}^{(p,q)}(z)/g_{n;j}^{(p,q)}(z)$, with $f_{n;j}^{(p,q)}(z)$ and $g_{n;j}^{(p,q)}(z)$ both polynomials. As we found from the equivalent differential equation~\eqref{diff-Pn}, $P_{n;j}^{(p,q)}(z)$ is analytic across the complex plane, and as such the solution has no poles. This implies that $g_{n;j}^{(p,q)}(z)$ is at most a complex constant. In particular, for the initial condition $P_{0;j}^{(p,q)}(z)$ in~\eqref{initial} we get $g_{0;j}^{(p,q)}(z)=1$. Using the latter and~\eqref{Pn} we obtain that $g_{1;j}^{(p,q)}(z)$ is a real constant, and so on for any higher iteration from~\eqref{Pn}. From this analysis, we conclude that $P_{n;j}^{(q,p)}(z)$ is a polynomial on the complex plane, and so it is on the real line $x\in\mathbb{R}$. 

Furthermore, the orthogonality relation~\eqref{inner-1} fulfilled by the eigensolutions of $H$ becomes handy to establish the orthogonality of the polynomials $P_{n;j}^{(p,q)}(x)$. In this form, it is straightforward to prove that $\{P_{n;1}^{(p,q)}(x)\}_{n=0}^{p}$ and $\{P_{m;2}^{(p,q)}(x)\}_{m=0}^{\infty}$ are orthogonal sets of polynomials with respect to the inner product
\begin{equation}
\int_{\mathbb{R}}dx [\mu^{(p,q)}(x)]^{2} P_{n;j}^{(p,q)}(x)P_{m;k}^{(p,q)}(x)=\delta_{n,m}\delta_{j,k} [\mathcal{N}_{n;j}^{(p,q)}]^{2} \, ,
\end{equation}
with the normalization constants given in~\eqref{N01}-\eqref{N02}. 

The exact form of the newly constructed polynomials $P_{n;j}^{(p,q)}(x)$ is depicted in Tab.~\ref{tab2} and Tab.~\ref{tab3} for the sequence $j=1$ and $j=2$, respectively, for several values of $q$, $p$, and $n$. The polynomials $P_{0;1}^{(p,q)}(x)=H_{p,2q}(x)$ and $P_{0;2}^{(p,q)}(x)=H_{p+1,2q+1}(x)$ are excluded from such table as those were presented in Tab.~\ref{tab1} already.

\begin{table}
	\centering
	\footnotesize{\begin{tabular}{cccp{12cm}}
			\hline
			$p$ & $q$  &n & $P_{n;1}^{(p,q)}(x)$ \\
			\hline
			2 &  1 & 1 & $-4 x^5-4 x^3-3 x$ \\
			2 &  1 & 2 & $8 x^6+12 x^4+18 x^2-9$ \\
			2 &  1 & $\geq 3$ & 0 \\
			\hline
			2 &  2 & 1 & $-16 x^9-96 x^7-216 x^5-120 x^3-45 x$  \\
			2 &  2 & 2 & $32 x^{10}+240 x^8+720 x^6+600 x^4+450 x^2-225$ \\
			2 &  2 & $\geq 3$ & 0 \\
			\hline
			3 &  1 & 1 & $-8 x^7+4 x^5-10 x^3-15 x$  \\
			3 &  1 & 2 & $16 x^8+40 x^4-15$  \\
			3 &  1 & 3 & $-16 x^9-72 x^5+135 x$  \\
			3 &  1 & $\geq 4$ & 0 \\
			\hline
	\end{tabular}}
	\caption{Polynomials $P^{(p,q)}_{n;j}(x)$ for the sequence $j=1$ computed through the three-term recurrence relation~\eqref{Pn}. The polynomials $P^{(p,q)}_{0;1}(x)=H_{p,2q}(x)$ serve as the initial condition.}
	\label{tab2}
\end{table}

\begin{table}
	\centering
	\footnotesize{\begin{tabular}{cccp{12cm}}
			\hline
			$p$ & $q$  &n & $P_{n;2}^{(p,q)}(x)$ \\
			\hline
			2 &  1 & 1 & $32 x^{10}-48 x^8+144 x^6-120 x^4-270 x^2+45 $ \\
			2 &  1 & 2 & $32 x^{11}-112 x^9+192 x^7-336 x^5-210 x^3+315 x $ \\
			2 &  1 & 3 & $64 x^{12}-384 x^{10}+720 x^8-1344 x^6+252 x^4+1512 x^2-189 $ \\
			\hline
			2 &  2 & 1 & $256 x^{16}+1536 x^{14}+6144 x^{12}+7296 x^{10}-4320 x^8+30240 x^6-50400 x^4-37800 x^2+4725 $  \\
			2 &  2 & 2 & $256 x^{17}+1024 x^{15}+3072 x^{13}-4608 x^{11}-21600 x^9+25920 x^7-90720 x^5+42525 x$ \\
			2 &  2 & 3 & $512 x^{18}+768 x^{16}-30720 x^{12}-54720 x^{10}+90720 x^8-285120 x^6+194400 x^4+182250 x^2-18225$ \\
			\hline
	\end{tabular}}
	\caption{Polynomials $P^{(p,q)}_{n;j}(x)$ for the sequence $j=2$ computed through the three-term recurrence relation~\eqref{Pn}. The polynomials $P^{(p,q)}_{0;2}(x)=H_{p+1,2q+1}(x)$ serve as the initial condition.}
	\label{tab3}
\end{table}

\section{Realization of the $su(1,1)$ and $su(2)$ algebras}
\label{sec:algebra}
Since the eigensolutions of $H$ decompose into a finite-dimensional and an infinite-dimensional sequence of polynomials, it is worth identifying the realization of any specific algebra associated with the ladder operators $A$ and $A^{\dagger}$ when acted on the appropriate sequence. As discussed earlier, the whole vector space $\mathcal{H}$ composes as the direct sum of the vector spaces $\mathfrak{h}_{1}$ and $\mathfrak{h}_{2}$ generated by each individual sequence. In this way, we may construct new sets of ladder operators out of $A$ and $A^{\dagger}$ so that the Lie algebras~\cite{Per88} $su(1,1)$ and $su(2)$ are recovered once acted on the appropriate vector space. This can be achieved by using the construction of $f$-oscillators~\cite{Man97} in terms of $A$ and $A^{\dagger}$. To this end, we introduce the new operators
\begin{equation}
\begin{cases} 
\widetilde{A}_{-}:=f(H)A \\ \widetilde{A}_{+}:=A^{\dagger}f(H) 
\end{cases} 
, \quad 
\begin{cases} 
\widetilde{B}_{-}:=g(H)A \\ \widetilde{B}_{+}:=A^{\dagger}g(H) 
\end{cases}
,
\label{f-osc}
\end{equation}
where $f(H)$ and $g(H)$ are functions of the Hamiltonian $H$ to be determined. More specifically, $f(H)$ is obtained by imposing that $\widetilde{A}_{\pm}$ fulfill the $su(2)$ commutation relations when acted on the elements of $\mathfrak{h}_{1}$. Similarly, $g(H)$ is determined by imposing that $\widetilde{B}_{\pm}$ fulfills the $su(1,1)$ commutation relations in $\mathfrak{h}_{2}$.

The $su(2)$ representation is thus recover by fixing $f(H)$ so that the relations
\begin{equation}
[\widetilde{A}_{-},\widetilde{A}_{+}]\phi^{(p,q)}_{n;1}=-2\widetilde{A}_{0}\phi^{(p,q)}_{n;1} \, , \quad [\widetilde{A}_{0},\widetilde{A}_{\pm}]\phi^{(p,q)}_{n;1}=\pm\widetilde{A}_{\pm}\phi^{(p,q)}_{n;1} \, , \quad \widetilde{A}_{0}:=\widetilde{N}_{A}-\mathfrak{a}_{0}\mathbb{I}_{\mathfrak{h}_{1}} \, ,
\label{su2}
\end{equation}
are met, with $\mathfrak{a}_{0}$ a real constant, together with $\widetilde{N}_{A}$ and $\mathbb{I}_{\mathfrak{h}_{1}}$ the number and identity operators in $\mathfrak{h}_{1}$, respectively. 

Similarly, for $su(1,1)$ we use
\begin{equation}
[\widetilde{B}_{-},\widetilde{B}_{+}]\phi^{(p,q)}_{n;2}=2\widetilde{B}_{0}\phi^{(p,q)}_{n;2} \, , \quad [\widetilde{B}_{0},\widetilde{B}_{\pm}]\phi^{(p,q)}_{n;2}=\pm\widetilde{B}_{\pm}\phi^{(p,q)}_{n;2} \, , \quad \widetilde{B}_{0}:=\widetilde{N}_{B}+\mathfrak{b}_{0}\mathbb{I}_{\mathfrak{h}_{2}} \, ,
\label{su11}
\end{equation}
where $\mathfrak{a}_{0}$ is another real constant, and $\widetilde{N}_{B}$ and $\mathbb{I}_{\mathfrak{h}_{2}}$ are the number and identity operators in $\mathfrak{h}_{2}$, respectively. 

Remark that the action of $f(H)$ and $g(H)$ on the elements of $\mathfrak{h}_{1}$ and $\mathfrak{h}_{2}$, respectively, take the form
\begin{equation}
f(H)\phi_{n;1}^{(p,q)}=f(2n)\phi_{n;1}^{(p,q)} \, , \quad g(H)\phi_{n;2}^{(p,q)}=g(2n+2p+4q+2)\phi_{n;2}^{(p,q)} \, .
\label{fgphi}
\end{equation}
This allows obtaining two finite difference equations for each case, from which the sets $\{\widetilde{A}_{\pm},\widetilde{A}_{0}\}$ and $\{\widetilde{B}_{\pm},\widetilde{B}_{0}\}$ can be uniquely determined. After some calculations we get
\begin{equation}
\begin{aligned}
& \left[f(2n-2)\right]^{2}2^{3}n(n-p-1)(n-p-2q-1)=-n^{2}-(2\mathfrak{a}_{0}-1)n+\mathfrak{a}_{1} \, , \\
& \left[g(2n+2p+4q)\right]^{2}2^{3}n(n+2q)(n+p+2q+1)=n^{2}+(2\mathfrak{b}_{0}-1)n+\mathfrak{b}_{1} \, ,
\end{aligned}
\label{fg-finite-diff}
\end{equation}
where $\mathfrak{a}_{1}$ and $\mathfrak{b}_{1}$ are integration constants. From~\eqref{fg-finite-diff}, the functions $f$ and $g$ are respectively fixed as
\begin{equation}
\left[f(2n)\right]^{2}=2^{-3}(p+2q-n)^{-1} \, , \quad \left[g(2n)\right]^{2}=2^{-3}(n+1)^{-1} \, ,
\label{fg}
\end{equation}
so that the action of $\{\widetilde{A}_{\pm},\widetilde{A}_{0}\}$ and $\{\widetilde{B}_{\pm},\widetilde{B}_{0}\}$ on the respective hierarchies of solutions $\{\phi^{(p,q)}_{n;1}\}_{n=0}^{p}$ and $\{\phi_{n;2}^{}(p,q)\}_{n=0}^{\infty}$ are
\begin{equation}
\begin{alignedat}{3}
&\widetilde{A}_{-}\phi_{n;1}^{(p,q)}=\sqrt{n(p+1-n)}\phi_{n-1;1}^{(p,q)} \, , \quad &&\widetilde{A}_{0}\phi_{n;1}^{(p,q)}=\left(n+\frac{p}{2}\right)\phi_{n;1}^{(p,q)} \, , \\
&\widetilde{B}_{-}\phi_{n;2}^{(p,q)}=\sqrt{n(n+2q)}\phi_{n-1;2}^{(p,q)} \, , \quad && \widetilde{B}_{0}\phi_{n;2}^{(p,q)}=\left(n+q+\frac{1}{2}\right)\phi_{n;2}^{(p,q)} \, . 
\end{alignedat}
\label{ABtilde}
\end{equation}

Notice that $\widetilde{A}_{-}$ and $\widetilde{A}_{+}$ are still annihilation and creation operators on $\mathcal{H}$. Nevertheless, the $su(2)$ realization is only valid in $\mathfrak{h}_{1}$. Similarly, the operators $\widetilde{B}_{-}$ and $\widetilde{B}_{+}$ realize the $su(1,1)$ relations only in $\mathfrak{h}_{2}$. 


\section{Conclusions}
The families of third-order shape-invariant potentials associated with the $-1/x$ and $-2x$ hierarchies of rational solutions to the fourth Painlev\'e equation have been identified in the most general form. Such identification has been made in terms of the generalized Hermite polynomials, which has allowed a quite general approach to the problem. Such polynomials have several handy properties to ensure the required conditions. For instance, the regularity of the resulting potential can be ensured by analyzing the zeros of $H_{p,q}$, which in turn also provide information about the corresponding zero-mode eigensolutions. This reveals that the set of eigenvalues decomposes into two disjoint sets, one of dimension $p+1$ and another one of infinite dimension. It was noticed that the general eigensolutions are composed of a weight function times a polynomial. This allows translating the spectral problem into the realm of orthogonal polynomials, for which we were able to determine the corresponding differential, and finite-difference equations fulfill for such polynomials. The latter appears in the form of a three-term recurrence relation, with the corresponding set of initial conditions written in terms of generalized Hermite polynomials, which is a helpful expression as one can compute any higher polynomial by simple linear recursions. This introduces a simplification compared to conventional approaches, where a higher-order nonlinear recurrence equation usually emerges. Our linear recurrence relation appears by exploiting the third-order ladder operators and by separating the general spectral problem into its two corresponding sequences. 

Coincidentally, it turns out that the ladder operators $A$ and $A^{\dagger}$ can be modified utilizing the approach of $f$-oscillators so that the new resulting operators fulfill the $su(1,1)$ and $su(2)$ algebraic relationships when acted on the finite- and infinite-dimensional space of solutions, respectively, while preserving the ladder operator structure in the general space of solutions. In this form, we have an alternative representation for such algebras in terms of the newly introduced polynomials $P_{m;j}^{(p,q)}(x)$. The latter can be straightforwardly implemented in the construction of coherent states, such as the $su(1,1)$ and $su(2)$ coherent states. However, this is beyond the scope of the present work and will be discussed somewhere else.


\section*{Acknowledgments}
I. Marquette was supported by Australian Research Council Future Fellowship FT180100099. K. Zelaya acknowledges the support from the project ``Physicists on the move II'' (KINE\'O II), Czech Republic, Grant No. CZ.02.2.69/0.0/0.0/18\_053/0017163; and the support of Consejo Nacional de Ciencia y Tecnolog\'ia, Mexico, grant number A1-S-24569. 



\end{document}